\documentclass{ws-ijmpe}
\begin{document}

\markboth{Christian Holm Christensen}{The ALICE Forward Multiplicity Detector}

\catchline{}{}{}{}{}
%%%%%%%%%%%%%%%%%%%%%%%%%%%%%%%%%%%%%%%%%%%%%%%%%%%%%%%%%%%%%%%%%%%%

\title{The ALICE Forward Multiplicity Detector}

\author{\underline{Christian Holm Christensen}, Jens J\o{}rgen
  Gaardh\o{}je, Kristj\'an Gulbrandsen, B\o{}rge Svane Nielsen,
  Carsten S\o{}gaard\\(For the ALICE Collaboration)}

\address{Niels Bohr Institute, University of Copenhagen, Blegdamsvej 17\\
Copenhagen, DK--2100, Denmark, cholm@nbi.dk}

\maketitle

\begin{history}
\received{(received date)}
\revised{(revised date)}
%\accepted{(Day Month Year)}
%\comby{(xxxxxxxxxx)}
\end{history}

\begin{abstract}
The ALICE Forward Multiplicity Detector (FMD) is a silicon strip
detector with 51,200 strips arranged in 5 rings, covering the range
$-3.4 < \eta < 5.1$.  It is placed around the beam pipe at
small angles to extend the charged particle acceptance of ALICE into the
forward regions, not covered by the central barrel detectors. 
\end{abstract}

\section{Motivation and purpose}

Experiences from RHIC have shown, that of interesting physics is
happening not only in the central rapidity region of A+A collisions,
but also at higher values of pseudo--rapidity.  In particular, the
BRAHMS collaboration has measured particle distributions and ratios,
high $p_\perp$ suppression, and spectra at very forward rapidity, and
obtained important information about the initial state of the colliding
nucleons (the colour glass
condensate)\cite{Arsene:2004ux,Arsene:2004fa}.

%% does not provide a measurement of particle species and momentum,
%% it does 

The ALICE Forward Multiplicity Detector\cite{unknown:2004mz} (FMD)
provides a high resolution charged particle multiplicity determination
at very forward rapidity, which allows ALICE to give a more complete
picture of the bulk properties of the interactions in A+A collisions.
The FMDs coverage does not extend all the way into the fragmentation
region in $\sqrt{s_{NN}}=5.5\,\mbox{TeV}$ Pb+Pb collisions, but the
extended overall shape of the charged particle multiplicity
distribution can help differentiate between different
scenarios\cite{Bjorken:1982qr,Landau:1953gs,Belenkij:1956cd,Back:2001ae,Bearden:2003fw}
for the A+A collisions.

Likewise, the measurement of anisotropic azimuthally flow
$v_2$\cite{Voloshin:1994mz} at RHIC has proved to be an extremely
useful tool in characterising the ultra--relativistic heavy ion
collisions\cite{Voloshin:1999gs,Back:2002gz,Back:2004mh,Adams:2005dq}.
The measurement of $v_2$ requires a good and reliable measurement of
the event planes inclination from the horizontal plane
\cite{Wang:2006xz}.

The ALICE Forward Multiplicity Detector provides measurements of the
charged particle multiplicity in the forward regions $-3.4 \leq \eta
\leq -1.7$ and $1.7 \leq \eta \leq 5.0$, measurement of the event
planes inclination, as well as an independent measurement of $v_2$.
Other kinds of charged particle multiplicity analysis, such as
fluctuations in particle number, can also be extended to cover larger
$\eta$ regions by using the FMD. 

\begin{figure}[htbp]
  \centering
  \includegraphics[keepaspectratio,width=.8\textwidth]{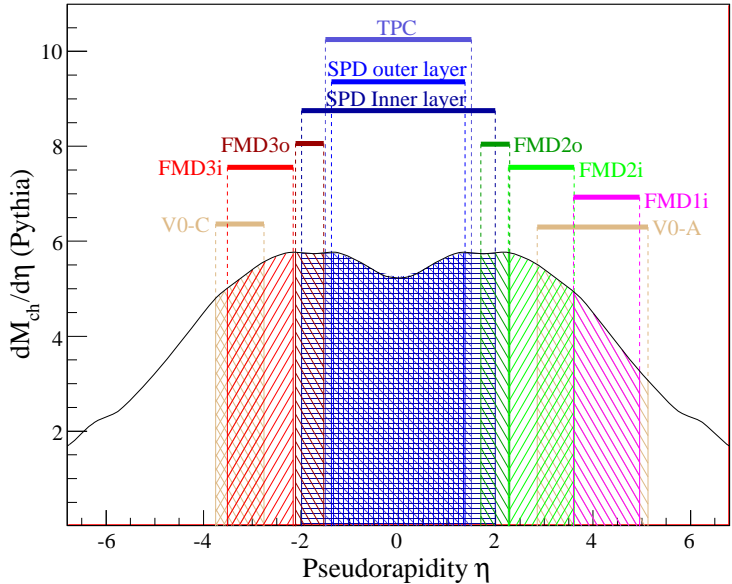}
  \caption[$\eta$ acceptance of the inner ALICE detectors.]{$\eta$
    acceptance of the inner ALICE detectors.  In the figure, the
    horizontal lines indicate the coverage of the first 2 layers (SPD
    inner and outer layer) of the Inner Tracking Systems, and the 5
    FMD rings.}
  \label{fig:alice:inner_acc}
\end{figure}

\section{The Detector} 

The FMD is a silicon strip detector of modest segmentation.  It is
made of 5 rings FMD1i, FMD2i, FMD2o, FMD3i, and FMD3o, placed around
the beam pipe.  The rings consists of 10 (for the inner rings, FMD1i,
FMD2i, and FMD3i) or 20 (for the outer rings FMD2o, FMD3o) hexagonal
silicon sensors cut out of $300\,\mu{}\mbox{m}$ thick silicon 6''
wafers.  Each sensor is azimuthal segmented into 2 sectors, and each
sector is segmented into strips at constant radii.  The segmentation
gives a total of 51,200 read-out channels (or strips).

\begin{figure}[htbp]
  \begin{center}
    \includegraphics[width=.7\textwidth]{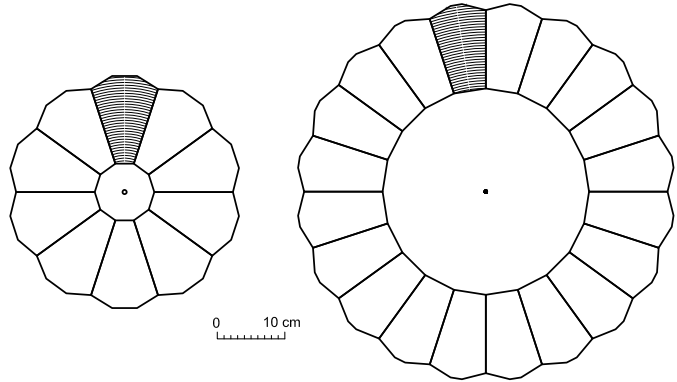}
    \caption[Assembly of inner and outer rings]{Assembly of an inner
      ring from 10 modules (left) and an outer ring from 20 modules
      (right). The size and shape of each module is determined by
      limitations imposed by the fabrication of sensors from 6~inch
      silicon wafers.}
    \label{fig:fmd:arrangment}
  \end{center}
\end{figure}

Fig.~\ref{fig:fmd:arrangment} shows the arrangement of sectors in the
inner and outer rings, and
\tablename~\ref{tab:fmd:table:stripdimension} gives the segmentation
of the various sub--detectors. Practically, the detector is assembled
in half rings to allow them to be mounted around thee LHC beam pipe.

\begin{table}[htbp]
  \begin{center}
    \caption{Physical dimensions of Si segments and strips.}
    \label{tab:fmd:table:stripdimension}
    \vglue0.2cm
    \begin{tabular}{|l|cc|cr@{\space--\space}l|r@{\space--\space}l|}
      \hline
      \textbf{Ring} &
      \textbf{Azimuthal}&
      \textbf{Radial} &
      $z$ &
      \multicolumn{2}{c|}{\textbf{$r$}} &
      \multicolumn{2}{c|}{\textbf{$\eta$}} \\
      &
      \textbf{sectors} &
      \textbf{strips} & 
      \textbf{[cm]} &
      \multicolumn{2}{c|}{\textbf{range [cm]}} &
      \multicolumn{2}{c|}{\textbf{coverage}} \\
      \hline
      FMD1 & 20& 512& 320  &  4.2& 17.2& 3.68&  5.03\\
      FMD2i& 20& 512&  83.4&  4.2& 17.2& 2.28&  3.68\\
      FMD2o& 40& 256&  75.2& 15.4& 28.4& 1.70&  2.29\\
      FMD3i& 20& 512& -75.2&  4.2& 17.2&-2.29& -1.70\\
      FMD3o& 40& 256& -83.4& 15.4& 28.4&-3.40& -2.01\\
      \hline
    \end{tabular}
  \end{center}
\end{table}

Mounted directly behind each sensor is a hybrid card with 8 (for the
inner rings) or 4 (for the outer rings) \texttt{VA1\_3} preamplifier
chips.  The \texttt{VA1\_3} belongs to the time--tried Viking family
of silicon preamplifier chips\cite{Toker:1993dd}.  The \texttt{VA1\_3}
is customised for the ALICE FMD to allow signals up to the equivalent
of 20 minimum ionising particles (MIP) impinging on the silicon
sensor.  Each preamplifier chip reads out 128 strips, and multiplexes
them into 1 differential read--out channel.

The sensor--hybrid modules are mounted on honeycomb plates for
support.  The modules are staggered in $z$ to minimise the dead zones
in the acceptance.  The failure rate of the modules has been found to
be $\ll 0.1\%$ -- that is, very few of the 51,200 strips have
problems.

A custom developed digitiser card (FMDD) is mounted on the back side
of each half--ring honeycomb support plates.  On the FMDD are three 16
channel ALTRO\cite{EsteveBosch:2003bj} ADC chips.  One read--out
channel of the \texttt{VA1\_3} preamplifier chip is connected to one
ALTRO channel.  As such, each ALTRO channel reads out 128 strips.  The
basic design of the FMDD is based on the TPC front end card, but
customised to the specific needs of the FMD.  For example, the FMD
requires additional trigger handling on the front-end not present on
the TPC front end card.

The ALTRO ADC is a custom chip developed for the ALICE TPC read-out
system.  It is a very advanced chip with many options for speeding up
the read-out and have the front-end do first-level data processing.
The ALTRO was chosen for the FMD, since it provides a fast parallel
read-out of many channels from the front end, is radiation tolerant,
and because it is highly customisable.

The FMDD has a Board Controller (BC), implemented in an FPGA, to
control the communication with the back-end of the read-out chain,
monitor the status of the electronics, control the preamplifiers, and
handle triggers.  The BC handles most of the tasks related to the
front end electronics.  It monitors that temperatures, voltages, and
currents are within acceptable values, and manages calibration runs.

The read-out of the sub-detectors FMD1, FMD2, and FMD3 is managed by a
Read--out Controller Unit\cite{EsteveBosch:2002xg} (RCU).  The RCU is
developed by the TPC collaboration, and controls the read out of the
ALTROs.  A special bus protocol has been developed to maximise the
throughput of data.  The RCU also provides means of communication for
the detector control system to control the front--end electronics via
a dedicated instruction memory.  Each RCU card is fitted with an
optical interface to transfer data to the data acquisition system.  A
daughter card with an embedded computer running Linux, provides the
interface to the outside world through standard TCP/IP communication.

\section{Test beam results}

The nearly complete system was tested with the
ASTRID\cite{Stensgaard:1988im} $e^-$ 650\,MeV parasitical beam.  The
sensors where put in a row behind each other (rather than in rings) to
allow maximum testing of the detector response to minimum ionising
particles.

\begin{figure}[htbp]
  \centering
  \includegraphics[keepaspectratio,width=.6\columnwidth]{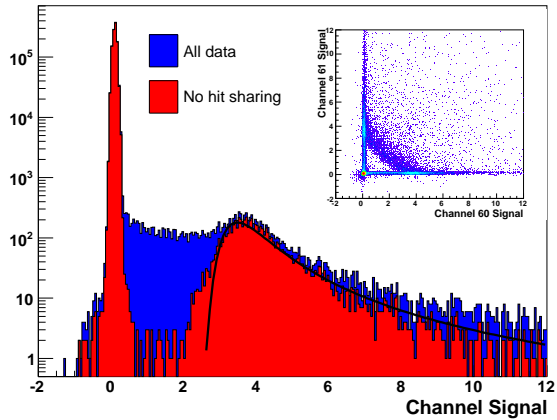}
  \caption[Laboratory test system data]{Laboratory test system data,
    before and after data is corrected for sharing (signal from one
    MIP shared of 2 or 3 channels). Insert shows anti--correlation of
    neighbouring strips.} 
  \label{fig:fmd:sharing}
\end{figure}

If the incident particle trajectory is not perpendicular to the plane
defined by the sensors, then a particle close to the strip boundaries
can deposit energy in more than one strip.  This gives rise to what is
known as \emph{sharing}.  It manifests itself as two or more adjacent
strips having signals over background less than the level
corresponding to a minimum ionising particle.  However, the sum of the
signals should corresponds to at least 1 MIP.
Fig.~\ref{fig:fmd:sharing} shows the distribution of signals before
and after cutting away the shared hits.  The insert shows the
correlation plot of recorded energy in adjacent strips.  The
anti-correlation seen in the insert is typical of sharing.

\begin{figure}[htbp]
  \centering
  \includegraphics[keepaspectratio,width=.9\columnwidth]{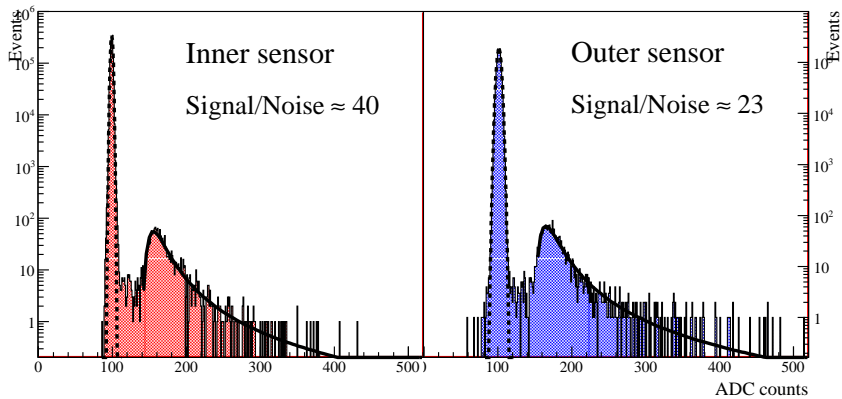}
  \caption[ADC spectrum for an inner and outer sensor]{ADC spectrum
    for an inner and outer sensor.  Disregarding the sharing, a
    signal--to--noise ratio of $\approx40:1$ for the inner
    type sensors  is seen.  For the outer type sensors a
    signal--to--noise ratio of $\approx23:1$ is seen. }  
  \label{fig:fmd:result} 
\end{figure}

Using the final read--out system, five sensors were put in a row in
the $e^-$ beam, and scintillators were placed before and after the
sensors and the coincidence between these was used as the trigger.
The resulting energy distribution is shown in
Fig.~\ref{fig:fmd:result}.  A signal--to--noise of $\approx 40:1$ for
the inner type sensors, and $\approx 23:1$ for the outer type sensors
was achieved.  This is well above the Technical Design
Report\cite{unknown:2004mz} requirement of a signal--to--noise ratio
of 10:1, and close to what can be expected from basic calculations.

\end{document}